\documentstyle[12pt]{article}
\catcode`@=11
\def\un#1{\relax\ifmmode\@@underline#1\else
        $\@@underline{\hbox{#1}}$\relax\fi}
\catcode`@=12

\def\a{\alpha}
\def\b{\beta}
\def\c{\chi}

\def\e{\epsilon}
\def\f{\phi}
\def\g{\gamma}
\def\h{\eta}

\def\j{\psi}
\def\k{\kappa}
\def\l{\lambda}
\def\m{\mu}
\def\n{\nu}

\def\p{\pi}
\def\q{\theta}
\def\r{\rho}

\def\t{\tau}

\def\F{\Phi}

\def\S{\Sigma}

\def\cd{{\cal D}}

\def\bo{{\raise-.5ex\hbox{\large$\Box$}}}               
\def\TH{{\raise.2ex\hbox{$\displaystyle \bigodot$}\mskip-4.7mu \llap H \;}}
\def\face{{\raise.2ex\hbox{$\displaystyle \bigodot$}\mskip-2.2mu \llap {$\ddot
        \smile$}}}                                      

\def\sp#1{{}^{#1}}                              
\def\leftrightarrowfill{$\mathsurround=0pt \mathord\leftarrow \mkern-6mu
        \cleaders\hbox{$\mkern-2mu \mathord- \mkern-2mu$}\hfill
        \mkern-6mu \mathord\rightarrow$}
\def\dvec#1{\vbox{\ialign{##\crcr
        \leftrightarrowfill\crcr\noalign{\kern-1pt\nointerlineskip}
        $\hfil\displaystyle{#1}\hfil$\crcr}}}           

\def\frac#1#2{{\textstyle{#1\over\vphantom2\smash{\raise.20ex
        \hbox{$\scriptstyle{#2}$}}}}}                   
\def\sfrac#1#2{{\vphantom1\smash{\lower.5ex\hbox{\small$#1$}}\over
        \vphantom1\smash{\raise.4ex\hbox{\small$#2$}}}} 
\def\bfrac#1#2{{\vphantom1\smash{\lower.5ex\hbox{$#1$}}\over
        \vphantom1\smash{\raise.3ex\hbox{$#2$}}}}       
\def\afrac#1#2{{\vphantom1\smash{\lower.5ex\hbox{$#1$}}\over#2}}    

\def\[{\lfloor{\hskip 0.35pt}\!\!\!\lceil}
\def\]{\rfloor{\hskip 0.35pt}\!\!\!\rceil}

\def\fracm#1#2{\hbox{\large{${\frac{{#1}}{{#2}}}$}}}
\def\half{{\fracm12}}

\def\un{\underline}

\def\low#1{{\raise -3pt\hbox{${\hskip 0.75pt}\!_{#1}$}}}

\newskip\humongous \humongous=0pt plus 1000pt minus 1000pt
\def\caja{\mathsurround=0pt}
\def\eqalign#1{\,\vcenter{\openup2\jot \caja
        \ialign{\strut \hfil$\displaystyle{##}$&$
        \displaystyle{{}##}$\hfil\crcr#1\crcr}}\,}
\newif\ifdtup

\def\ref#1{$\sp{#1)}$}

\def\pl#1#2#3{Phys.~Lett.~{\bf {#1}B} (19{#2}) #3}
\def\np#1#2#3{Nucl.~Phys.~{\bf B{#1}} (19{#2}) #3}

\def\mpl#1#2#3{Mod.~Phys.~Lett.~{\bf A{#1}} (19{#2}) #3}

\def\sfo{S\!F\!O}
\def\ico{I\!C\!O}

\begin{document}


\begin{titlepage}

\noindent
ITP--UH--13/96 \\
hep-th/9607106 \hfill July 1996\\

\vskip 0.6cm

\begin{center}

{\Large\bf The Self-Dual Critical $N{=}2$ String$^{+*}$} \\

\vskip 1.5cm

{\large Olaf Lechtenfeld}

\vskip 0.6cm

{\it Institut f\"ur Theoretische Physik, Universit\"at Hannover}\\
{\it Appelstra\ss{}e 2, 30167 Hannover, Germany}\\
{http://www.itp.uni-hannover.de/\~{}lechtenf/}\\

\vskip 2cm
{\bf Abstract}
\end{center}

\begin{quote}
\hspace{\parindent}
I review the covariant quantization of the closed fermionic string with 
(2,2) extended world-sheet supersymmetry on ${\bf R}^{2,2}$.
Results on $n$-point scattering amplitudes are presented, for tree- and
one-loop world-sheets with arbitrary Maxwell instanton number.  I~elaborate 
the connection between Maxwell moduli, spectral flow, and instantons.
It is argued that the latter serve to extend the Lorentz symmetry from
$U(1,1)$ to $SO(2,2)$ by undoing the choice of spacetime complex structure.
\end{quote}

\vfill

\textwidth 6.5truein
\hrule width 5.cm

{\small
\noindent ${}^+$
Talk given at the International Conference 
`Problems of Quantum Field Theory' in \\
\phantom{${}^+$} Alushta, Crimea, Ukraine, 16 March 1996 \\
\noindent ${}^*\;$
Supported by the `Volkswagen Foundation'
and the `Deutsche Forschungsgemeinschaft'
}

\eject
\end{titlepage}

\newpage
\hfuzz=10pt

\noindent {\bf 1. Introduction}\\
In this talk I am going to report on recent results \cite{bkl,kl,bu,bi,bl}
obtained in collaboration with Jan Bischoff, a PhD student, 
and Sergei Ketov, a senior postdoc at Hannover, Germany.
Fermionic strings with extended world-sheet supersymmetry are much
less understood than the usual bosonic or superstrings, although
their discovery now dates back twenty years. For a review, see~\cite{ma,ov}.
Responsible for this gap is the fact that the physical spectrum of
extended fermionic strings is not rich enough to be of immediate
relevance for particle phenomenology.
Although strings with $N{=}2$ world-sheet supersymmetry are critical
in four dimensions, the spacetime signature must be (2,2), and the
excitation spectrum contains only a single massless scalar field.
Thus, such strings are particular point particle theories in disguise
and describe self-dual gravitational and gauge fields~\cite{ov}.
On the other hand, 
these peculiarities attract the interest of mathematical physicists.
For example, $N{=}2$ strings should teach us something 
about the quantization of integrable systems.
Furthermore, they can be embedded into a twisted $N{=}4$ string
which hints at a topological interpretation~\cite{bv2,ber,ov2}.
Finally, $N{=}2$ heterotic strings form a building block in the recently
proposed F-theory~\cite{F} unifying superstring compactification schemes, and
they should play some role in the emerging picture of string-string duality.

It is well-known that the low-energy effective actions (LEEAs) of $N{=}0$
(bosonic) open and closed strings is Yang-Mills and gravity 
(plus other fields), respectively, in $1{+}25$ spacetime dimensions.
In the $N{=}1$ case, the five cases of supergravity and/or super-Yang-Mills
in $1{+}9$ spacetime dimensions emerge in the low-energy limit.
In contrast, open and closed $N{=}2$ strings do not give rise to
spacetime supersymmetry. Astoundingly, their LEEAs represent
{\it self-dual\/} Yang-Mills and {\it self-dual\/} gravity in
$2{+}2$ spacetime dimensions, respectively. The coupling of both from 
open-string loops or heterotic versions, however, has not been fully clarified.

\smallskip
\noindent {\bf 2. Covariant Quantization}\\
The Polyakov formulation of the $(2,2)$ fermionic string involves
complexified world-sheet scalar string coordinates $X^\m$ and their 
complexified world-sheet Majorana spinor partners $\j^\m$, with $\m=0,1$, 
coupled to the $N{=}2$ supergravity multiplet consisting of the real
world-sheet metric~$h_{\a\b}$, complex gravitino~$\c_\a$, 
and real Maxwell field~$A_\a$ (graviphoton), with $\a,\b=0,1$.
The Brink-Schwarz action~\cite{bsa} possesses $N{=}2$ super-diffeomorphism and
$N{=}2$ super-Weyl invariance on the world-sheet as well as global 
$U(1,1)\times{\bf Z}_2$ target-space symmetry.
It is important to note that only the world-sheet spinors $\j^\m$ and $\c_\a$
are charged under the local $U(1)$ Maxwell symmetry, 
whereas global $U(1,1)$ Lorentz transformations
affect only the matter fields $X^\m$ and~$\j^\m$.

After $N{=}2$ superconformal gauge fixing of the Brink-Schwarz action,
the local residual gauge transformations decompose chirally.
The left- and right-moving parts each comprise an $N{=}2$ superconformal
algebra whose generators $T$, $G^\pm$, and $J$ act as constraints
in the quantized theory.
Proper gauge fixing in the euclidean path integral expression of 
a closed string amplitude induces Faddeev-Popov determinants, 
which are rewritten in terms of real conformal $(b_{\a\b},c^\a)$, 
complex (spinorial) superconformal $(\b_\a,\g)$, 
and real Maxwell $(\tilde b_\a,\tilde c)$ ghost systems.
The anti-ghost zero modes are projected out by appropriate anti-ghost insertions
into the path integral; eventual ghost zero modes are associated with
global residual gauge symmetries and have to be absorbed by vertex operators.
The implementation of the first-class constraints in the ghost-extended
version is most elegantly performed in the BRST framework.

Cancellation of conformal and chiral anomalies requires
the flat target space to be complex two-dimensional. 
In the following I shall specialize to the simplest case of ${\bf C}^{1,1}$.
Then, world-sheet bosons must be single-valued, whereas non-trivial
monodromies are still allowed for world-sheet fermions $f\in\{\j,\c,\b,\g\}$. 
The latter being complex (which I denote by a $\pm$ superscript), one may have
$$
f^\pm\ \longrightarrow\ e^{\pm2\pi i\r}\ f^\pm\qquad{\rm with}\quad\r\in[0,1]
\eqno(1)$$
for the transportation around a non-trivial loop.
Thus, the theory sports a continuous set of sectors parametrized by~$\r$ and
interpolating between the usual NS ($\r{=}\half$) and R ($\r{=}0$) sectors.
This phenomenon is well-known as {\it spectral flow\/} 
in $N{=}2$ superconformal models.

The computation of a closed string $n$-point amplitude
requires an integration over all $N{=}2$ superconformal structures, 
or supermoduli, of $2d$ $n$-punctured orientable $N{=}2$ supermanifolds~$s\S_n$. 
The universal supermoduli space decomposes into a sum 
indexed by the genus~$g\in{\bf Z}_+$ and the instanton number~$c\in{\bf Z}$.
This reflects the topological classification of the tangent and principal 
$U(1)$ bundles over ordinary compact Riemann surfaces~$\S$ through
$$
\frac{1}{2\p}\int_\S R \ =\ 2-2g \qquad\quad{\rm and}\qquad\quad
\frac{1}{2\p}\int_\S F \ =\ c
\eqno(2) $$
where $R$ and $F$ are the curvature two-forms of the spin and Maxwell 
connections of $N{=}2$ world-sheet supergravity, respectively.

The supermoduli spaces have a natural complex structure and are parametrized 
by the {\it meromorphic\/} differentials of order $2$, $\frac32$, and $1$,
on~$\S$, with Maxwell charges of $0$, ${\pm}\half$, and $0$, respectively.
Locally, each supermoduli space is a direct sum of 
metric, fermionic, and Maxwell moduli spaces, with complex dimensions of 
$3g{-}3{+}n$, $2g{-}2{\pm}c{+}n$, and $g{-}1{+}n$, respectively.
Metric and fermionic moduli may be treated like in the $(1,1)$ fermionic string. 
Explicitly integrating out the fermionic moduli leads to additional insertions
into the path-integral measure, which combine with certain anti-ghost insertions
to picture-changing operators.
If $|c|>2g{-}2{+}n$, fermionic zero modes can no longer be soaked up in the
path integral, which then must vanish.
Hence, only a finite number of instanton sectors contribute at a given genus.

In contrast, the Maxwell moduli are specific to the $N{=}2$ string.
They parametrize the space of flat connections or harmonic one-forms~$H$
on the $n$-punctured world-sheet~$\S_n$ of genus~$g$. 
Since $H=m+\bar m$ where $m$ is a meromorphic one-form
with a single pole at each puncture~$p_\ell$,
a basis is provided by the normalized first abelian differentials 
and the normalized third abelian differentials.
Thus, the Maxwell moduli space is a complex $(g{+}n{-}1)$-dimensional torus,
parametrized by $\oint H$ for the basic homology cycles 
around ($a_i$) and through $(b_i$) the handles as well as 
around ($c_\ell$) and between ($d_\ell$) the punctures.

A closed string $n$-point amplitude with external momenta 
$k_1,\ldots,k_n\in{\bf C}^{1,1}$ has the following structure,
$$
A^n(k_1,\ldots,k_n)\ =\
\sum_{g=0}^\infty\ \k^{2g-2+n} \sum_{c=-2g+2-n}^{2g-2+n}\!\!\l^{-c} \;
\int\!d\t\; A^n_{g,c}(k_1,\ldots,k_n|\t)
\eqno(3) $$
where $\k$ and $\l$ are the gravitational and Maxwell string couplings,
respectively, and $\t$ collectively denotes the metric and Maxwell moduli.
The fixed-moduli correlation function reads
$$
A^n_{g,c}(k_1,\ldots,k_n|\t)\ =\ 
\int\!\cd[X\j bc\b\g\tilde b\tilde c]\;{\rm AZI}\cdot{\rm PCO}\,
\prod_{\ell=1}^n V_\ell(k_\ell)\; e^{iS[X\ldots|\t]}
\eqno(4) $$
with anti-ghost zero mode and picture-changing insertions AZI and PCO,
vertex operators~$V_\ell$ and the gauge-fixed action~$S$.

To enumerate the spectrum of physical excitations of $N{=}2$ strings,
one needs first to determine their BRST cohomology
$H(Q)=ker Q/im Q =\{|{\rm phys}\rangle\}$
and then to list additional identifications among physical states,
e.g. by picture-changing or spectral flow.
This investigation has not been completed, but partial results
are very suggestive~\cite{bkl,bien,lp}.
Namely, for a given lightlike momentum $k^{\pm\m}$ 
one finds a single bosonic scalar state $|k^{\pm\m}\rangle$,
but no states whatsoever at any massive level!
This agrees with the naive expectation that 2+2 dimensions leave no
room for transverse vibrations.
In essence, the field theory of $N{=}2$ strings is really just the
field theory of a scalar particle, described by the spacetime field~$\F$!
Therefore, one should expect the interaction of $N{=}2$ strings to be
fully given by the exact LEEA.
Accordingly, the task is to compute all scattering amplitudes for~$\F$
at tree-, loop-, and multi-instanton-level in $N{=}2$ strings.

\smallskip
\noindent {\bf 3. Amplitudes}\\
Let me present the results~\cite{bu,bi} of direct computations of closed 
$N{=}2$ on-shell string amplitudes (at fixed metric and Maxwell moduli).
The lightlike momenta will be denoted by $k_\ell^{\pm\m}$,
with $\ell{=}1,\ldots,n$, $\ \m{=}0,1$, and the $\pm$ superscript indicating 
the charge $q{=}\pm\half$ under the $U(1)$ factor of the $U(1,1)$ Lorentz group.
With the 2d invariant tensors $\h_{\m\n}$ and $\e_{\m\n}$ 
the on-shell condition reads
$\h_{\m\n}k_\ell^{+\m}k_\ell^{-\n}=0$.
Momentum conservation, $\sum_\ell k_\ell^{\pm\m}=0$, will be assumed.
We introduce the following three antisymmetric bilinears,
$$
c^0_{\ell m}\ :=\ \h_{\m\n}\,(k_\ell^{+\m}k_m^{-\n}-k_\ell^{-\m}k_m^{+\n})
\qquad{\rm and}\qquad
c^\pm_{\ell m}\ :=\ \e_{\m\n}\, k_\ell^{\pm\m}k_m^{\pm\n} \quad,
\eqno(5) $$
of which the first one is fully $U(1,1)$ invariant while the other two
are only inert under $SU(1,1)$ as their charge index indicates.

For less than three external legs,
$$
A^0\ =\ 0\quad,\qquad A^1\ =\ 0\quad,\qquad A^2\ =\ 1\quad,
\eqno(6) $$
including loop and instanton corrections.
The tree-level three-point function is
$$
A^3_{g=0}\ =\ \l^{+1}(c^-_{23})^2\ +\ \l^0(c^0_{23})^2\ +\ \l^{-1}(c^+_{23})^2
\eqno(7) $$
with the observation that Lorentz invariance requires the coupling $\l$ to
acquire a $U(1)$ charge of two units!
Using the standard Mandelstam variables $s$, $t$, and $u$,
the first non-trivial amplitude is
$$
\eqalign{
A^4_{g=0}\ \propto\ 
   &\l^{+2} [c^-_{12}c^-_{34}t + c^-_{23}c^-_{41}s - stu]^2 \
+\  \l^{+1} [c^-_{12}c^0_{34}t + c^-_{23}c^0_{41}s - stu]^2 \cr
&\qquad\qquad+\ \l^0 [c^0_{12}c^0_{34}t + c^0_{23}c^0_{41}s - stu]^2 \cr
+\ &\l^{-1} [c^+_{12}c^0_{34}t + c^+_{23}c^0_{41}s - stu]^2 \
+\  \l^{-2} [c^+_{12}c^+_{34}t + c^+_{23}c^+_{41}s - stu]^2 \cr 
=\ \ &0 \cr }
\eqno(8) $$
due to the very special kinematics in 2+2 dimensions!
It is believed that $A^{n\ge4}=0$ to all orders, 
so that string duality is consistent with the conjectured complete absence 
of massive poles in intermediate channels.

At the one-loop level, we have computed~\cite{bi,bgi}
$$
A^3_{g=1}\ \propto\ \Bigl( A^3_{g=0} \Bigr)^3
\eqno(9) $$
with the help of 
$c^+_{\ell m} c^-_{\ell m} + c^0_{\ell m} c^0_{\ell m} = 0$,
leading me to the interesting conjecture that
$$
A^3_g\ \propto\ \Bigl( A^3_{g=0} \Bigr)^{2g+1} \quad.
\eqno(10) $$

\noindent {\bf 4. Spectral Flow and Instantons}\\
Since moduli are integrated over, shifting their values cannot affect
the final amplitude.
It does, however, have an effect on the integrand, $A^n_{g,c}$.
The Maxwell moduli couple to the gauge-fixed action as
$$
S[X\ldots|\t]\ \sim\ \int_\S H\wedge*J
\eqno(11) $$
so that $H\to H+h$ changes the action by
$$
\int_\S h\wedge*J\ =\ \sum_{i=1}^g \Bigl(
\oint_{a_i}\!h\oint_{b_i}\!*J - \oint_{b_i}\!h\oint_{a_i}\!*J \Bigr)
+\ \sum_{\ell=1}^n \oint_{c_\ell}\!h\int_{p_0}^{p_\ell}\!\!*J
\eqno(12) $$
where I used the harmonicity of $h$ and $d*J{=}0$ and introduced 
a reference point~$p_0$.
Clearly, my shift has induced twists $\exp\{2\pi i\q\int *J\}$
around the standard homology cycles as well as around
the punctures, with $\q\in[0,1]$.
Dependence on $p_0$ drops out because $\sum_\ell{\rm res}_{p_\ell}h=0$.

When integrating over the Maxwell moduli, I could fix some~$H$
and let $h$ sweep over the complex torus.
During this course, the twists extend the already existing 
{\it sum\/} over spin structures to an {\it integral\/} over spin structures.
Integration over matter and ghost fields finally yields Jacobi theta functions, 
with the prescription to integrate over their characteristics~\cite{bi}.
This is in line with the phenomenon of spectral flow.
Indeed, a twist around a puncture at~$z$ may be absorbed into a
redefined (twisted) vertex operator,
$$
V(z)\ \longrightarrow\ V^{(\q)}(z)\ :=\ e^{2\pi i\q\int_{z_0}^z *J}\,V(z)\quad,
\eqno(13) $$
which can be seen to create a spectrally flowed state from the vacuum.
Alternatively, conjugating the generators of the $N{=}2$ superconformal
algebra by the spectral-flow operator
$$
\sfo(\q)\ :=\ \exp\{2\pi i\q\int_{z_0}^z *J\}
\eqno(14) $$
explicitly realizes the spectral flow endomorphism of the algebra.

Furthermore, $\sfo$ is BRST-closed but only its zero mode is not
BRST-exact, so that its position in an amplitude does not matter.
Hence, amplitudes are twist-invariant,
$\langle V_1^{(\q_1)}\ldots V_n^{(\q_n)}\rangle =
\langle V_1 \ldots V_n \rangle$,
as long as the total twist vanishes, $\sum_\ell\q_\ell{=}0$.
Bosonizing the local $U(1)$ current $*J=d\f$ with a linear combination 
$\f=\f_m+\f_{gh}$ of compact (matter and ghost) bosons, 
$$
\sfo(\q)\ =\ \exp\Bigl\{2\pi i\q\ \bigl[\f(z)-\f(z_0)\bigr]\Bigr\}
\ =\ e^{-2\pi i\q\f(z_0)}\ \exp\Bigl\{2\pi i\q\ \f(z)\Bigr\}
\eqno(15) $$
provides me with a local operator $\exp\{2\pi i\q\f\}$, 
modulo zero-mode ambiguities.
The two factors in eq.~(15) are separately neutral under the 
{\it local\/} $U(1)$, but carry opposite {\it global\/} $U(1)$ charges.
\goodbreak

Interestingly, spectral flow is related to Maxwell instantons on the
world-sheet.
Flowing around full circle, i.e. choosing $\q{=}1$, one may define
an instanton-creation operator
$$
\ico\ :=\ \l\ \sfo(\q{=}1)\ =\ \l\ e^{-2\pi i\f(z_0)}\ \exp\{2\pi i\f(z)\}
\eqno(16) $$
which implements a singular gauge transformation 
$A\to A+d\ln z$
and changes the world-sheet instanton number via
$\ico|c\rangle = |c{+}1\rangle$.
I can now hide the zero-mode or reference-point ambiguity by
simply declaring
$$
\l\ \equiv\ e^{2\pi i\f(z_0)} 
\eqno(17) $$
which gives $\ico$ and $\l$ (!) a global $U(1)$ charge of $q{=}2$.
Then, amplitudes with different instanton backgrounds are related through
$$
\langle V_1\ldots V_n\ \rangle_c \ =\
\langle V_1\ldots V_n\ (\ico)^c\ \rangle_{c=0}\ =\
\l^c\ \langle V_1^{(\q_1)}\ldots V_n^{(\q_n)}\rangle_{c=0}
\eqno(18) $$
with a total twist of $\sum_\ell\q_\ell=c$. 
This matches the $\l$ dependence in eq.~(3).

\smallskip
\noindent {\bf 5. Instantons as Lorentz Generators}\\
Finally, I propose a common solution of two related problems.
First, I have shown that $A^3_{c\ne0}$ is only invariant under the
$SU(1,1)$ {\it subgroup\/} of the spacetime Lorentz group.
Second, I should even like to {\it extend\/} the Lorentz group from 
$U(1,1)\simeq U(1)\times SU(1,1)$ to $SO(2,2)\simeq SU(1,1)\times SU(1,1)$,
as is proper for a real (2+2)-dimensional spacetime and necessary for
the existence of {\it spacetime\/} spinors.
Surprisingly, my instanton creation and annihilation operators point to
a solution. 
Factorizing into a matter and ghost part, $\ico=\ico_m \ico_{gh}$,
the first factor is a genuine current and provides me with new charges,
$M_+=\oint\ico_m$ and $M_-=\oint\ico_m^{-1}$.
Together with the $U(1)$ Lorentz generator $M_3=\oint d\f_m$, they create 
the {\it second\/} $SU(1,1)$ factor of the extended $SO(2,2)$ Lorentz group!

It is not difficult to check that $M_+$ and $M_-$ act on amplitudes by simply 
raising and lowering the $U(1)$ charge~$q$ of two $c^q_{\ell m}$
kinematical factors by one unit each. For $n{=}3$,
the range of instanton backgrounds in eq.~(3), $|c|\le2g{+}1$,
is nicely matched by the allowed range of $U(1)$ charges, $|q|\le2(2g{+}1)$,
from eqs.~(10) and~(7).
These charge assignments are also compatible with amplitude factorization.

To conclude, the extended $SO(2,2)$ Lorentz invariance of $N{=}2$ string
amplitudes is not manifest but realized in a hidden way.
The inevitable choice of a complex structure in ${\bf R}^{2,2}$ 
selects some $U(1,1)$ subgroup.
In essence, {\it world-sheet\/} Maxwell instantons rotate around this subgroup
and, with it, the {\it spacetime\/} complex structure. 
It is therefore natural to view the Maxwell coupling~$\l$ as a 
{\it coordinate\/} parametrizing the moduli space of complex structures,
which requires it to appropriately transform under part of the Lorentz group.
This seems to connect well with the results of~\cite{bv2,ov2}
and asks for a twistor-like formulation.

\end{document}